\documentclass[11pt,a4paper]{article}
\usepackage[T1]{fontenc}
\usepackage[latin1]{inputenc}
\usepackage{amsmath,amssymb,amsthm}
\def\be{\begin{equation}}
\def\ee{\end{equation}}
\def\bea{\begin{eqnarray}}
\def\eea{\end{eqnarray}}

\newcommand{\rmd}{\mathrm{d}}

\newcommand{\ket}[1]{|\kern.3ex#1\kern.3ex\rangle}
\newcommand{\bra}[1]{\langle\kern.3ex #1 \kern.3ex|}

\begin{document}

\begin{center}
{\Large \bf  Random Aharonov-Bohm vortices and some exact
families of integrals: Part III}\\[0.5cm]

{\large  \bf St\'ephane Ouvry}\footnote{stephane.ouvry@u-psud.fr}\\[0.1cm]
Universit\'e Paris-Sud, Laboratoire de Physique Th\'eorique et Mod\`eles
Statistiques\footnote{Unit\'e Mixte de Recherche CNRS-Paris Sud, UMR 8626}\\
91405 Orsay, France
\\[0.2cm]
\end{center}

\begin{abstract}
As a sequel to  \cite{moi} and  \cite{mash}, I present some recent progress  on   Bessel integrals $\int_0^{\infty}{\rmd u}\;
uK_0(u)^{n}$, $\int_0^{\infty}{\rmd u}\;
u^{3}K_0(u)^{n}$, ... where  the power of the integration variable  is odd  and where $n$, the Bessel weight,  is a positive integer. Some of these integrals for weights $n=3$ and $n=4$ are known to be intimately related to the zeta numbers $\zeta(2)$ and $\zeta(3)$.
 Starting from a Feynman diagram inspired representation  in terms of  $n$ dimensional multiple integrals on an infinite domain,  one shows how to partially integrate  to     $n-2$ dimensional multiple integrals on a finite domain. In this process the Bessel integrals are shown to be periods\footnote{ Periods are defined in  \cite{Konsev} as "values of absolutely convergent integrals of rational functions with rational coefficients over domains in $R^n$ given by polynomial inequalities with rational coefficients".}. Interestingly enough, these "reduced" multiple integrals can be considered  in parallel  with some simple integral representations of $\zeta$ numbers.  One also generalizes the construction of \cite{mash} on a particular sum of double nested Bessel integrals to a whole family of double nested integrals. Finally a strong PSLQ  numerical evidence is  shown to support  a surprisingly simple  expression  of $\zeta(5)$ as a linear combination with rational coefficients of   Bessel integrals of weight $n= 8$.

\end{abstract}

\section{Introduction: the  quantum mechanics random magnetic impurity model and Bessel integrals}\label{arxiv}

The random magnetic impurity model \cite{nous} describes a quantum particle in a plane coupled to a random distribution of  Aharonov-Bohm fluxes  perpendicular to the plane. It was introduced having in mind the Integer Quantum Hall effect. A perturbative expansion of the partition function of the  model in the  disorder coupling constant $\alpha$ (the Aharonov-Bohm flux  expressed in  unit of the quantum of flux) for $2$ impurities, i.e. at second order  in the impurity density $\rho$,   lead us to consider Feynmann diagrams with maximal impurity line crossing  at order $\rho^2\alpha^4$ \cite{nous} and $\rho^2\alpha^6$ \cite{private}.

On the one hand, these Feymann diagrams, which reduce after momenta integrations to  multiple integrals on intermediate temperatures,
 were shown to rewrite in terms of simple integrals \cite{nousbis} 
\begin{align} I_{\rho^2\alpha^4} =  \int_0^{\infty}u\,K_0(u)^2(uK_1(u))^2 \rmd u\end{align} 
 and double nested integrals \cite{private} 
 \begin{align}I_{\rho^2\alpha^6} = & 8  \int_0^{\infty}\rmd u \,u \,K_0(u)^2(uK_1(u))^2
\int_0^{u}\rmd x\,xK_1(x)I_1(x)K_0(x)^2 \nonumber \\
&-4  \int_0^{\infty}\rmd u \, u K_0(u)(uK_1(u))\left( uK_1(u)I_0(u)-uI_1(u)K_0(u)\right)
\int_u^{\infty}\rmd x\, x\,K_0(x)^2K_1(x)^2 \nonumber\\ 
&+  \int_0^{\infty}u\,K_0(u)^4(uK_1(u))^2 \rmd u \label{wellbis}
\end{align}
on products of  modified Bessel functions  $K_{\nu}$ and $I_{\nu}$, with $\nu=0,1$.

On the other hand, one could show \cite {nousbis} by direct integration  
 \be\label{first0}\int_0^{\infty} u K_0(u)^4 \rmd u={2^3-1\over 8}\zeta(3)\ee 
and by integration by part that 
\be \label{first}I_{\rho^2\alpha^4}=\int_0^{\infty}u\,K_0(u)^2(uK_1(u))^2 \rmd u\ee 
is a linear combination with rational coefficients\footnote{ Linear combination  with rational coefficients   means  here that there exist three positive or negative integers $a, b$ and $c$  such that \be \nonumber a\int_0^{\infty}u\,K_0(u)^2(uK_1(u))^2 \rmd u+b\int_0^{\infty} u K_0(u)^4 \rmd u+c=0\ee } of $1$ and $\int_0^{\infty} u K_0(u)^4 \rmd u$, i.e. of $1$ and $(2^3-1)\zeta(3)$.
Similarly one obtained \cite{private,moi} by direct integration\footnote{See the Appendix  for the derivations of (\ref{first0}) and (\ref{ffirst}).} 
{\be \label{ffirst}\int_{0}^{\infty}\rmd u\,  
uI_0(u)K_{0}(u)^3={2^2-1\over 8}\zeta(2)\ee} 
and by integration by part that
\be\label{firstbis} \int_0^{\infty}\rmd u\,uK_0(u)(uK_1(u))^2I_0(u)\ee
and 
\be\label{firstbis1} \int_0^{\infty}\rmd u\,uK_0(u)^2uK_1(u)uI_1(u)\ee
are linear combinations of $1$ and $\int_{0}^{\infty}\rmd u\, u 
I_0(u)K_{0}(u)^3$, i.e. of $1$ and $(2^2-1)\zeta(2)$.
Likewise,  by integration by part \cite{mash}  
\be\label{rel}\int_{0}^{\infty}u \, K_0(u)^4(uK_1(u))^2\,\rmd u =
{2\over 15}\int_{0}^{\infty}u \, K_0(u)^6\,\rmd u
-{1\over 5}\int_{0}^{\infty}u^3 \, K_0(u)^6\,\rmd u \ee 

It was then natural to argue  in \cite{mash} that the sum of double nested integrals (\ref{wellbis})  might also rewrite as a linear combination with rational coefficients of simple integrals on product of Bessel functions of weight 6 -defined as the total power of  Bessel functions- and of $\zeta$ numbers of weigth $6-1= 5$, like $\zeta(5)$, or below. The "counting rule" inferred from (\ref{first0}, \ref{first}, \ref{ffirst}, \ref{firstbis}, \ref{firstbis1}, \ref{rel}) is that an  integration $\int_0^{\infty} {\rmd u}\;u^n(\quad)$ with $n$ odd diminishes the  Bessel weight by one, so $\int_0^{\infty}u\,K_0(u)^4(uK_1(u))^2$ is like  $\zeta(5)$,  that a Bessel function $I_{\nu}$  has  no weight, so $\int_{0}^{\infty}\rmd u\,  
uI_0(u)K_{0}(u)^3$  is like $\zeta(2)$ and $\int_0^{\infty}\rmd u \,u^3 \,K_0(u)^2K_1(u)^2
\int_0^{u}\rmd x\,xK_1(x)I_1(x)K_0(x)^2$  like $\zeta(3)\zeta(2)$ i.e. like $\zeta(5)$, that  in turn  is like $\int_0^{\infty}u\,K_0(u)^6$
or    $\int_0^{\infty}u^3\,K_0(u)^6$  -for an explanation why only odd powers of $u$ and no higher power appear here see \cite{mash}  (see also  subsection (\ref{remark})). Indeed in \cite{mash} a numerical PSLQ \cite{PSLQ} search  gave 
\be\label{PSLQ}
{I_{\rho^2\alpha^6}}=_{\rm PSLQ}{1\over 30}\int_{0}^{\infty}u \, K_0(u)^6\,\rmd u
+ {1\over 20}\int_{0}^{\infty}u^3 K_0(u)^6\,\rmd u -{2^5-1\over 160}\,\zeta(5)
\ee
(the $2^5-1$ factor multiplying $\zeta(5)$ has to be viewed in parallel with $2^3-1$ multiplying $\zeta(3)$ in (\ref{first0}) and  $2^2-1$ multiplying $\zeta(2)$ in (\ref{firstbis})). Note that in (\ref{PSLQ}), as in the sequel,
  an  identity obtained from a numerical PSLQ search  is labelled as $=_{\rm PSLQ}$.

It follows that the perturbative expansion of the random magnetic impurity model has Bessel integrals for building blocks. Note that  Bessel integrals  also appear in other physical contexts such as, for example,  the Ising model \cite{Kreimerbis}. 

In \cite{mash} it was argued that all integrals in the family 
\be\label{set}\int_0^{\infty}u^{p+1} K_0(u)^{n-j} (uK_1(u))^{j} \, \rmd u\quad\quad p \;{\rm even}\;\ge 0\ee  can  be expressed as  linear combinations with rational coefficients of a finite set of Bessel integrals $\int_0^{\infty}{\rmd u}\;
uK_0(u)^{n}$, $\int_0^{\infty}{\rmd u}\;
u^{3}K_0(u)^{n}$, ... , the latter  constituting a basis for the former.  In the present work one concentrates on different  issues such as integrating the Bessel integrals.


\section{ Integrating $\int_{0}^{\infty}\rmd u\,  u
K_{0}(u)^n$}\label{integrating}

\small{As alluded to in the Introduction,   Feyman diagrams  momenta integrations  
 lead  to multiple integrals on  differences of consecutive intermediate temperatures. In the random magnetic impurity model, these multiple integrals take a form such that Bessel integrals 
can quite generally be     representated in a similar way: 
change  variable $u=2\sqrt{t}$ and use the integral representation 
 \be\label{represent} \int_0^{\infty}\rmd a\,   a^{\nu -1}e^{ -a-{t\over a}}
=2K_{\nu}(u)({u\over 2})^{\nu}\ee
so that
\begin{align}&\int_{0}^{\infty}\rmd u\,  u
K_{0}(u)^n\nonumber\\
=&\int_{0}^{\infty}2\rmd t   \int_{0}^{\infty}{\rmd a_1\over 2} \ldots \int_{0}^{\infty}{\rmd a_n\over 2}{1\over a_1 a_2\ldots a_n}e^{ -a_1-{t\over a_1}}e^{ -a_2-{t\over a_2}}\ldots e^{ -a_n-{t\over a_n}}\nonumber\\
=&2  \int_{0}^{\infty}{\rmd a_1\over 2} \ldots \int_{0}^{\infty}{\rmd a_n\over 2}{1\over a_1 a_2\ldots a_n}{1\over{1\over a_1}+{1\over a_2}+\ldots+{1\over a_n}}e^{ -a_1-a_2-\ldots-a_n}\nonumber\\
=&{1\over 2^{n-1}}  \int_{0}^{\infty}{\rmd a_1} \ldots \int_{0}^{\infty}{\rmd a_n}{1\over a_1 a_2\ldots a_{n-1}+ a_2 a_3\ldots a_{n}+\ldots+ a_{n} a_1\ldots a_{n-2}}e^{ -(a_1+a_2+\ldots+a_n)}\label{feyn}\end{align}

I am now going to show how to integrate the $n$ dimensional multiple integral (\ref{feyn}) to a $n-2$ multiple integral, which in turn will share some similarity  with a well-known integral representation of $\zeta$ numbers. In the process the $n-1$ to $n-2$  variables
reduction will  be a non trivial example of a multidimensional integration leading to a simple and elegant expression.  Note that it has so far  not possible to integrate further along the same lines.

\subsection{\bf $a_{n}$ integration} 
Introduce the variable $\beta$ to rewrite $e^{ -(a_1+a_2+\ldots+a_n)}=\int_{0}^{\infty}{\rmd} \beta\,e^{-\beta}\delta(\beta-(a_1+a_2+\ldots+a_n))$, then integrate over $a_n$
\begin{align}&\int_{0}^{\infty}\rmd u\,  u
K_{0}(u)^n\nonumber\\   
=&{1\over 2^{n-1}} \int_{0}^{\infty}{\rmd \beta} e^{-\beta } \int_{0}^{\beta}{\rmd a_1} \int_{0}^{\beta-a_1}{\rmd a_2}\ldots \int_{0}^{\beta-a_1-\ldots-a_{n-2}}{\rmd a_{n-1}}\nonumber\\ 
&{1\over a_1 a_2\ldots a_{n-1}+ (\beta-(a_1+a_2+\ldots+a_{n-1}))(a_2a_3 \ldots a_{n-1}+a_3a_4 \ldots a_{n-1}a_1+\ldots+  a_1\ldots a_{n-2} )}\end{align}
Change variables $a_i\to a'_i={ a_i/ \beta}$ (notation $a'_i\to a_i$)
{so that the $\beta$ integration becomes trivial and finally}  
\begin{align}&\int_{0}^{\infty}\rmd u\,  u
K_{0}(u)^n={1\over 2^{n-1}} \int_{0}^{1}{\rmd a_1} \int_{0}^{1-a_1}{\rmd a_2}\ldots \int_{0}^{1-a_1-\ldots-a_{n-2}}{\rmd a_{n-1}}\nonumber\\ 
&{1\over a_1 a_2\ldots a_{n-1}+ (1-(a_1+a_2+\ldots+a_{n-1}))(a_2a_3 \ldots a_{n-1}+a_3a_4 \ldots a_{n-1}a_1+\ldots+  a_1a_2\ldots a_{n-2} )}\end{align}
With  the notations  
 
$u_{n}=a_1 +a_2+\ldots+ a_{n}$ 

$v_{n}= a_2a_3 \ldots a_{n}+a_3a_4 \ldots a_{n}a_1+\ldots+ a_{n}a_1 a_2\ldots a_{n-2}+a_1a_2\ldots a_{n-1}$ 

$w_{n}=a_1 a_2\ldots a_{n}$

\noindent one has shown that
\begin{align}\int_{0}^{\infty}\rmd u\,  u
K_{0}(u)^n&={1\over 2^{n-1}}  \int_{0}^{\infty}{\rmd a_1} \int_{0}^{\infty}{\rmd a_2}\ldots \int_{0}^{\infty}{\rmd a_n}{1\over v_n}e^{ -u_n}\nonumber\\
&={1\over 2^{n-1}} \int_{0}^{1}{\rmd a_1} \int_{0}^{1-u_1}{\rmd a_2}\ldots \int_{0}^{1-u_{n-2}}{\rmd a_{n-1}}{1\over w_{n-1}+ (1-u_{n-1}) v_{n-1}}\label{etoui}
\end{align}
 \subsection{\bf $a_{n-1}$ integration}
Use 
$$ u_{n-1}=a_{n-1}+u_{n-2}, \quad v_{n-1}=a_{n-1}v_{n-2}+w_{n-2}, \quad w_{n-1}=a_{n-1}w_{n-2}$$
so that
\begin{align}
&{1\over w_{n-1}+ (1-u_{n-1}) v_{n-1}}
= {1\over w_{n-2}(1-u_{n-2})+a_{n-1}v_{n-2}(1-u_{n-2})-a_{n-1}^2 v_{n-2}}\nonumber\\
&={1\over -v_{n-2}(a_{n-1}-a_{n-1}^+)(a_{n-1}-a_{n-1}^-)} ={1\over- v_{n-2}(a_{n-1}^{+}-a_{n-1}^{-})}({1\over a_{n-1}-a_{n-1}^{+}}-{1\over a_{n-1}-a_{n-1}^{-} })\end{align}
where $$a_{n-1}^{\pm}={-v_{n-2}(1-u_{n-2})\pm\sqrt{v_{n-2}^2(1-u_{n-2})^2+4w_{n-2}v_{n-2}(1-u_{n-2})}\over -2v_{n-2}}$$

\noindent One finds
\begin{align} 
&\int_{0}^{\infty}\rmd u\,  u
K_{0}(u)^n={1\over 2^{n-1}} \int_{0}^{1}{\rmd a_1} \int_{0}^{1-u_1}{\rmd a_2}\ldots \int_{0}^{1-u_{n-3}}{\rmd a_{n-2}}\nonumber\\ 
&{1\over \sqrt{v_{n-2}^2(1-u_{n-2})^2+4w_{n-2}v_{n-2}(1-u_{n-2})}}\log|{a_{n-1}^{-}\over a_{n-1}^{+}}{1-u_{n-2}-a_{n-1}^{+}\over 1-u_{n-2}-a_{n-1}^{-}}|\end{align}
where the integrand rewrites as
\begin{align}{2\over \sqrt{v_{n-2}^2(1-u_{n-2})^2+4w_{n-2}v_{n-2}(1-u_{n-2})}}\log|{a_{n-1}^{-}\over a_{n-1}^{+}}|={2\over (1-u_{n-2})v_{n-2}}X\log{1+X\over 1-X}\end{align}
with 
\be X=\sqrt{1\over 1+{4w_{n-2}\over (1-u_{n-2})v_{n-2}}}\label{0change}\ee
At this point a seemingly complicated -square root- expression of the remaining $n-2$ variables appears. However  the 
change of variables 
\be \label{change} a_i\to x_i=X {a_i\over u_{n-2}}\ee 
so that when $0<a_i<1$ then  $1>X>0\Rightarrow 1>x_i>0$ will give a more tractable form for the integrand. The    Jacobian  is
$$ {(u_{n-2})^{n-2}\over X^{n-3}|\sum_{i=1}^{n-2}a_i\partial_i X|}$$
Use homogeneity relations
\begin{align}
\sum_{i=1}^{n-2}a_i\partial_i u_{n-2}&=u_{n-2}\nonumber\\
\sum_{i=1}^{n-2}a_i\partial_i v_{n-2}&=(n-3)v_{n-2}\nonumber\\ 
\sum_{i=1}^{n-2}a_i\partial_i w_{n-2}&=(n-2)w_{n-2}\end{align}
so that
$$\sum_{i=1}^{n-2}a_i\partial_i X=-{1\over 2}X(1-X^2){1\over 1-u_{n-2}}$$
and
\begin{align}\int_{0}^{\infty}\rmd u\,  u
K_{0}(u)^n=&{1\over 2^{n-1}} \int_{0}^{1}{\rmd x_1} \int_{0}^{1-x_1}{\rmd x_2}\ldots \int_{0}^{1-x_1-\ldots - x_{n-3}}{\rmd x_{n-2}}\log{1+X\over 1-X}\nonumber\\
&{2\over (1-u_{n-2})v_{n-2}}X{(u_{n-2})^{n-2}\over X^{n-3}|-{1\over 2}X(1-X^2){1\over 1-u_{n-2}}| }\nonumber\\
=&{1\over 2^{n-1}} \int_{0}^{1}{\rmd x_1} \int_{0}^{1-x_1}{\rmd x_2}\ldots \int_{0}^{1-x_1-\ldots - x_{n-3}}{\rmd x_{n-2}}\log{1+X\over 1-X}\nonumber\\
&{4\over v_{n-2}}{(u_{n-2})^{n-2}\over X^{n-3}(1-X^2) }\end{align} 
From (\ref{0change})  and (\ref{change})  it follows
\begin{align}X =&\; x_1+x_2+\ldots+x_{n-2}\nonumber\\w_{n-2} X^{n-2}=& \;(u_{n-2})^{n-2} x_1x_2\ldots x_{n-2}\nonumber\\ {u_{n-2}v_{n-2}\over w_{n-2}X}=&\;{1\over x_1}+{1\over x_2}+\ldots+{1\over x_{n-2}}
\nonumber\\{1\over X^2}-1=& \;{4w_{n-2}\over(1-u_{n-2})v_{n-2}}\label{ouaip} \end{align}
so that  finally  $\int_{0}^{\infty}\rmd u\,  u
K_{0}(u)^n$ takes the desired simple form
\begin{align}\label{nicenice}&\int_{0}^{\infty}\rmd u\,  u
K_{0}(u)^n ={1\over 2^{n-1}} \int_{0}^{1}{\rmd x_1} \int_{0}^{1-x_1}{\rmd x_2}\ldots \int_{0}^{1-x_1-\ldots - x_{n-3}}{\rmd x_{n-2}}\log{1+x_1+x_2+\ldots+x_{n-2}\over 1-(x_1+x_2+\ldots+x_{n-2})}\nonumber\\
&{4\over 4 (x_1+\ldots+x_{n-2})x_1\ldots x_{n-2}+(1-(x_1+\ldots+x_{n-2})^2)(x_2x_3\ldots x_{n-2}+\ldots+x_1x_2\ldots x_{n-3})}
\end{align} 
This  procedure generalises to $\int_{0}^{\infty}\rmd u\,  u^p
K_{0}(u)^n$ with $p$ odd: for example for $p=3$ one obtains similarly
 \begin{align}&\int_{0}^{\infty}\rmd u\,  u^3
K_{0}(u)^n={1\over 2^{n-3}} \int_{0}^{1}{\rmd a_1} \int_{0}^{1-a_1}{\rmd a_2}\ldots \int_{0}^{1-a_1-\ldots-a_{n-2}}{\rmd a_{n-1}}\nonumber\\
&{a_1a_2\ldots a_{n-1}(1- (a_1 +a_2+\ldots+ a_{n-1}))\over (a_1 a_2\ldots a_{n-1}+ (1-(a_1+a_2+\ldots+a_{n-1}))(a_2a_3 \ldots a_{n-1}+a_3a_4 \ldots a_{n-1}a_1+\ldots+  a_1a_2\ldots a_{n-2} ))^2}\nonumber\\
&={1\over 2^{n-3}} \int_{0}^{1}{\rmd x_1} \int_{0}^{1-x_1}{\rmd x_2}\ldots \int_{0}^{1-x_1-\ldots - x_{n-3}}{\rmd x_{n-2}}\nonumber\\
&\left({1+(x_1+x_2+\ldots+x_{n-2})^2\over 2(x_1+x_2+\ldots+x_{n-2})}\log{1+x_1+x_2+\ldots+x_{n-2}\over 1-(x_1+x_2+\ldots+x_{n-2})}-1\right)\nonumber\\ 
&{4x_1\ldots x_{n-2}\left(1-(x_1+x_2+\ldots+x_{n-2})^2\right)\over \bigg(4 (x_1+\ldots+x_{n-2})x_1\ldots x_{n-2}+(1-(x_1+\ldots+x_{n-2})^2)(x_2x_3\ldots x_{n-2}+\ldots+x_1x_2\ldots x_{n-3})\bigg)^2}\label{cela}\end{align} 

\subsection{Some remarks}\label{remark}
 
(\ref{nicenice}) and (\ref{cela}) indicate that $\int_{0}^{\infty}{\rmd u}\;u K_0(u)^n$ and $\int_{0}^{\infty}{\rmd u}\;u^3 K_0(u)^n$ are periods.
This is the case in general for  $\int_{0}^{\infty}{\rmd u}\;u^p K_0(u)^n$ with $p$ odd.
Note that  it is crucial  the power $p$ of the integration variable $u$  be odd  to  obtain  rational functions of the $n-2$ remaining variables $x_1,\ldots,x_{n-2}$ -if not square root  would appear in the process.

Looking  at (\ref{nicenice}), it  has been  found that $\int_{0}^{\infty}\rmd u\,  u
K_{0}(u)^n$ rewrites as a $n-2$ integral on the finite domain $0\le x_1+x_2+\ldots x_{n-2}\le 1$ with, for integrand, $\log{1+x_1+x_2+\ldots+x_{n-2}\over 1-(x_1+x_2+\ldots+x_{n-2})}$  multiplied by a rational function.  Interestingly enough, this structure shares some similarities with the well-known  rewriting  of $\zeta$ numbers in terms of a simple integral on the domain $0\le x\le 1$   with for integrand $(\log{1+x\over 1-x})^{n-1}$ multiplied by $1/x$
\be{2^n-1\over 2^{n-1}}\zeta(n)=\sum_{k=1}^{\infty}{1\over k^n}+(-1)^{k-1}{1\over k^n}={1\over 2}\int_0^{\infty}{\rmd x}x^{n-1}({1\over e^x-1}+{1\over e^x+1})={1\over (n-1)!}\int_0^1{{\rmd x}}(\log{1+x\over 1-x})^{n-1}{1\over x}\label{zetaint}\ee
 One can argue in (\ref{nicenice}) that  the logarithm has been linearized in place of  appearing in (\ref{zetaint}) at  a power $n-1$, but the price to be  paid is    a multiple integral on $n-2$ variables with the  variable $x$ in (\ref{zetaint}) replaced in the logarithm by the sum $x_1+x_2+\ldots+x_{n-2}$ and in the  factor $1/x$  by a more involved polynomial  of degree  $n-1$. 
 
 In the simple cases $n=3$ and $n=4$, one can indeed integrate (\ref{nicenice}) completely to simple expressions  in terms of $\zeta(2)$ and $\zeta(3)$ numbers (see (\ref{first0})   and (\ref{moncoco}, \ref{cb}) in the Appendix).

\subsection{A summary}

Finally if in (\ref{nicenice}) and (\ref{cela}) one changes notations $x_i\to a_i$  one has  shown
%
\begin{align} &2^{n-1}\int_{0}^{\infty}{\rmd u}\;u K_0(u)^n=\int_{0}^{\infty}{\rmd a_1}\ldots \int_{0}^{\infty}{\rmd a_{n-2}}\int_{0}^{\infty}{\rmd a_{n-1}}\int_{0}^{\infty}{\rmd a_{n}}\quad{1\over v_n}\quad e^{-u_{n}}\nonumber\\
&= \int_{0}^{1}{\rmd a_1} \int_{0}^{1-u_1}{\rmd a_2}\ldots \int_{0}^{1-u_{n-3}}{\rmd a_{n-2}}\int_{0}^{1-u_{n-2}}{\rmd a_{n-1}}\quad{1\over w_{n-1}+ (1-u_{n-1})v_{n-1}}\label{OK}\\
&=  \int_{0}^{1}{\rmd a_1}\int_{0}^{1-u_1}{\rmd a_2}\ldots \int_{0}^{1-u_{n-3}}{\rmd a_{n-2}}\quad\log{1+u_{n-2}\over 1-u_{n-2}}\quad{4\over 4 u_{n-2}w_{n-2}+(1-u_{n-2}^2)v_{n-2}}\label{jean}\end{align}
and
\begin{align}&2^{n-3}\int_{0}^{\infty}{\rmd u}\;u^3 K_0(u)^n= \int_{0}^{\infty}{\rmd a_1}\ldots \int_{0}^{\infty}{\rmd a_{n-2}}\int_{0}^{\infty}{\rmd a_{n-1}}\int_{0}^{\infty}{\rmd a_{n}}\quad{w_n\over v_n^2}\quad e^{-u_{n}}\nonumber\\ 
&= \int_{0}^{1}{\rmd a_1} \int_{0}^{1-u_1}{\rmd a_2}\ldots \int_{0}^{1-u_{n-3}}{\rmd a_{n-2}}\int_{0}^{1-u_{n-2}}{\rmd a_{n-1}}\quad{w_{n-1}(1-u_{n-1})\over (w_{n-1}+ (1-u_{n-1})v_{n-1})^2}\label{OKK}\\ 
&= \int_{0}^{1}{\rmd a_1}\int_{0}^{1-u_1}{\rmd a_2}\ldots \int_{0}^{1-u_{n-3}}{\rmd a_{n-2}}\left({1+u_{n-2}^2\over2u_{n-2} }\log{1+u_{n-2}\over 1-u_{n-2}}-1\right){4w_{n-2}(1-u_{n-2}^2)\over( 4 u_{n-2}w_{n-2}+(1-u_{n-2}^2)v_{n-2})^2}\nonumber\end{align}
As a remark one can proceed similarly with $\int_{0}^{\infty}\rmd u\,  u
I_0(u)K_{0}(u)^n$:
using again the integral representation (\ref{represent}) and ($u=2\sqrt{t}$)
\be I_0(u)=\sum_{k=0}^{\infty}{t^k\over (k!)^2}\ee
one can integrate over $t$ using 
\be \int_0^{\infty} \rmd t\; t^ke^{-t x}={k!\over x^{k+1}}\ee
to obtain 
\begin{align} &2^{n-1}\int_{0}^{\infty}{\rmd u}\;u I_0(u)K_0(u)^n=\int_{0}^{\infty}{\rmd a_1}\ldots \int_{0}^{\infty}{\rmd a_{n-2}}\int_{0}^{\infty}{\rmd a_{n-1}}\int_{0}^{\infty}{\rmd a_{n}}\quad{1\over v_n}\quad e^{\displaystyle -u_{n}+{w_n\over v_n}}\nonumber
\end{align}
Introducing as above the variable $\beta$, integrating over $a_n$ and then trivially over $\beta$ one finally obtains
\begin{align} &2^{n-1}\int_{0}^{\infty}{\rmd u}\;u I_0(u) K_0(u)^n\nonumber\\
&= \int_{0}^{1}{\rmd a_1} \int_{0}^{1-u_1}{\rmd a_2}\ldots \int_{0}^{1-u_{n-3}}{\rmd a_{n-2}}\int_{0}^{1-u_{n-2}}{\rmd a_{n-1}}\quad{1\over w_{n-1}u_{n-1}+ (1-u_{n-1})v_{n-1}}\label{OKKK}\end{align}
a result  to be compared to (\ref{etoui}). 
One could push the integration a step further   to obtain an  expression  in terms of $u_{n-2}, v_{n-2}$ and $w_{n-2}$ but somehow  more involved than (\ref{jean}) since its denominator would contain a square root.

}
\section{ Rewriting $I_{\rho^2\alpha^6}$ and a family of double nested Bessel integrals}\label{rewriting}
The fact that the double nested integrals in (\ref{wellbis})  can be  reexpressed in (\ref{PSLQ}) as a linear combination with rational coefficients of simple Bessel integrals  with the right weight -here weight $6$- and of $\zeta(5)$, fits well in the  "Bessel integral  $\to$ zeta number" mapping \cite{mash}. If
  a product  $f$ of Bessel functions   is, like in (\ref{first0}, \ref{first}, \ref{ffirst}, \ref{firstbis}, \ref{firstbis1}), mapped by simple integration on a $\zeta$ number denoted by ${\tilde \zeta}(f)$
\[f\to \int_0^{\infty}f(u)\rmd u ={\tilde\zeta}(f)\] 
then for a pair of such  products $f,g$  the mapping by  double "nested" integration
\[f,g\to \int_0^{\infty}f(u)\rmd u \int_0^u g(x)\rmd x={\tilde \zeta}(f,g)\]
on a polyzeta number denoted by ${\tilde \zeta}(f,g)$ 
makes sense, since, because of \[\int_0^{\infty}f(u)\rmd u \int_0^u g(x)\rmd x =
\int_0^{\infty}f(u)\rmd u \int_0^{\infty}
g(x) \rmd x -\int_0^{\infty}g(u)\rmd u \int_0^u f(x)\rmd x, \] 
one has
\be\label{polyfun}{\tilde\zeta}(f,g)={\tilde\zeta}(f){\tilde\zeta}(g)-{\tilde\zeta}(g,f)\ee
in analogy with
\be\label{poly}\zeta(p,q)=
\zeta(p)\zeta(q)-\zeta(p+q)-\zeta(q,p)\ee
for the standard polyzeta $\zeta(p,q)=\sum_{n>m}{1\over n^p}{1\over m^q}$ - if  $\zeta(p,q)$ would be defined as $\sum_{n>m}{1\over n^p}{1\over m^q}+{1\over 2}\zeta(p+q)$ then (\ref{poly}) would take the form  (\ref{polyfun}).

Coming back to (\ref{wellbis},\ref{PSLQ}) use \[\int_0^{\infty}g(u)\rmd u \int_u^{\infty} f(x)\rmd x =\int_0^{\infty}f(u)\rmd u \int_0^u g(x)\rmd x\] to rewrite $I_{\rho^2\alpha^6}$ as
\begin{align} \label{wellter}
I_{\rho^2\alpha^6} =  8 & \int_0^{\infty}\rmd u \,u \,K_0(u)^2(uK_1(u))^2
\int_0^{u}\rmd x\,xK_1(x)I_1(x)K_0(x)^2 \nonumber \\
-4&\int_0^{\infty}\rmd u\, u\,K_0(u)^2K_1(u)^2 \int_0^{u}\rmd x\, x K_0(x)xK_1(x)\big(xK_1(x)I_0(x)-xK_0(x)I_1(x)\big)
\nonumber\\ 
+&\int_0^{\infty}u\,K_0(u)^4(uK_1(u))^2 \rmd u 
\end{align}
Next use $xI_1(x)K_0(x)+xI_0(x)K_1(x)=1 $ so that 
\begin{align} \label{wellterter}
I_{\rho^2\alpha^6} =  8 &\int_0^{\infty}\rmd u \,u^3 \,K_0(u)^2K_1(u)^2
\int_0^{u}\rmd x\,xK_1(x)I_1(x)K_0(x)^2 \nonumber \\
 +8 &\int_0^{\infty}\rmd u \,u \,K_0(u)^2K_1(u)^2
\int_0^{u}\rmd x\,x^3K_1(x)I_1(x)K_0(x)^2 \nonumber \\
 -4&\int_0^{\infty}\rmd u\, u\,K_0(u)^2K_1(u)^2 \int_0^{u}\rmd x\, x^2 K_0(x)K_1(x)
\nonumber\\ 
+&\int_0^{\infty}u\,K_0(u)^4(uK_1(u))^2 \rmd u 
\end{align}
One has for the next to last  term in (\ref{wellterter})
  \begin{align} \label{wellquar}
 -4\int_0^{\infty}\rmd u\, u\,K_0(u)^2K_1(u)^2 \int_0^{u}\rmd x\, x^2 K_0(x)K_1(x) =2 &\int_0^{\infty}\, u\,K_0(u)^2K_1(u)^2 \big((u K_1(u))^2-1\big)\rmd u \nonumber\\
 =-{2\over 3}&\int_0^{\infty} u\, K_1(u) \,\big((u K_1(u))^2-1\big)\rmd K_0(u)^3\nonumber \\
={2\over 3}&\int_0^{\infty}   K_0(u)^3\,\rmd (u K_1(u))\big((u K_1(u))^2-1\big) \nonumber\\
={2\over 3}&\int_0^{\infty} u K_0(u)^4  \rmd u\, -2\int_0^{\infty}u\,K_0(u)^4(uK_1(u))^2 \rmd u
\end{align}
so that
\begin{align} \label{wellquarquar}
I_{\rho^2\alpha^6} = 8 &\int_0^{\infty}\rmd u \,u^3 \,K_0(u)^2K_1(u)^2
\int_0^{u}\rmd x\,xK_1(x)I_1(x)K_0(x)^2 \nonumber \\
+8 &\int_0^{\infty}\rmd u \,u \,K_0(u)^2K_1(u)^2
\int_0^{u}\rmd x\,x^3K_1(x)I_1(x)K_0(x)^2 \nonumber \\
+{2\over 3}&\int_0^{\infty} u K_0(u)^4  \rmd u
\nonumber\\ 
-&\int_0^{\infty}u\,K_0(u)^4(uK_1(u))^2 \rmd u 
\end{align}

Defining  
\[f_n(x) = x^n K_0(x)^2 K_1(x)^2\quad {\rm and} \quad g_n(x) = x^n  K_1(x) I_1(x)K_0(x)^2\]
\noindent equations (\ref{first}, \ref{rel},   \ref{wellquarquar}) imply that 
\begin{align} \label{nice}{\tilde \zeta}(f_3,g_1)+{\tilde \zeta}(f_1,g_3)=&{1\over 48}\int_{0}^{\infty}u \, K_0(u)^6\,\rmd u
- {3\over 160}\int_{0}^{\infty}u^3 K_0(u)^6\,\rmd u\nonumber\\&- {7\over 96}\,\zeta(3)-{31\over 1280}\,\zeta(5)
\end{align}
as confirmed by a direct  PSLQ check.  

So the   meaning of (\ref{PSLQ}) when rewritten as (\ref{nice}) might be that it is the symmetric form  ${\tilde \zeta}(f_3,g_1)+{\tilde \zeta}(f_1,g_3)$ which is a linear combination with rational coefficients of simple Bessel integrals of weigth $6$ and $\zeta$ numbers like $\zeta(5)$ and below. This suggests to look at other  symmetric sums of double nested integrals sharing this property. A PSLQ search confirms that
${\tilde \zeta}(f_5,g_1)+{\tilde \zeta}(f_1,g_5)$ belongs indeed to this category 
\begin{align} {\tilde \zeta}(f_5,g_1)+{\tilde \zeta}(f_1,g_5)&=_{PSLQ}{211\over 11520}\int_{0}^{\infty}u \, K_0(u)^6\,\rmd u
+ {3953\over 23040}\int_{0}^{\infty}u^3 K_0(u)^6\,\rmd u\nonumber\\&+{11\over 9216}- {1\over 9}\,\zeta(3)-{93\over 5120}\,\zeta(5)
\end{align}
as well as, for example,  ${\tilde \zeta}(f_7,g_1)+{\tilde \zeta}(f_1,g_7)$
  \begin{align} {\tilde \zeta}(f_7,g_1)+{\tilde \zeta}(f_1,g_7)&=_{PSLQ}{108731\over 1728000}\int_{0}^{\infty}u \, K_0(u)^6\,\rmd u
+{4256617\over 3456000}\int_{0}^{\infty}u^3 K_0(u)^6\,\rmd u\nonumber\\&+{27877\over 
  460800}- {8\over 15}\,\zeta(3)-{279
  \over 5120}\,\zeta(5)
\end{align}
  and ${\tilde \zeta}(f_3,g_5)+{\tilde \zeta}(f_5,g_3)$
\begin{align} {\tilde \zeta}(f_3,g_5)+{\tilde \zeta}(f_5,g_3)&=_{PSLQ}-{28921\over 691200}\int_{0}^{\infty}u \, K_0(u)^6\,\rmd u
+ {1151533\over 1382400}\int_{0}^{\infty}u^3 K_0(u)^6\,\rmd u\nonumber\\&+{14653\over 184320}+ {25\over 192}\,\zeta(3)+{279\over 20480}\,\zeta(5)
\end{align}

  One infers that  for any two odd positive integers $n,m$ one should have that ${\tilde \zeta}(f_n,g_m)+{\tilde \zeta}(f_m,g_n)$ can be rewritten as a linear combination of  $1,\; \int_{0}^{\infty}u \, K_0(u)^6\,\rmd u,\;\int_{0}^{\infty}u^3 \, K_0(u)^6\,\rmd u$, $\zeta(3)$ and $\zeta(5)$. Finding the coefficients of the linear combination can be in principle achieved by generalizing the integration by part procedures  \cite{ moi, mash, nous, nousbis} for simple integrals  to double nested integrals.

\section{Conclusion : Why looking at Bessel integrals?}

 In Section (\ref{arxiv}) some arguments (and examples) were given of why Bessel integrals are expected to be related to $\zeta$ numbers.  

On the one hand, one  has found  the surprisingly simple PSLQ identity
\be\label{final} \zeta(5)=_{\rm PSLQ}{1\over 77}\int_{0}^{\infty}u \, K_0(u)^8\,\rmd u
-{72\over 77}\int_{0}^{\infty}u^3 \, K_0(u)^8\,\rmd u 
\ee  
It would  certainly be rewarding to look for  a sytematics, if any,  behind, for example,  (\ref{first0}), (\ref{final}) and (\ref{moncoco}).

On the other hand,  one has tried, to this aim, to integrate the Bessel integrals starting from  Feynman diagram inspired multidimensional integrals on an infinite domain. In the process on has been able to reduce these integrals  via a non trivial multidimensional integration to  $n-2$ dimensional integrals on a finite domain, whose structure are reminiscent of  well-known simple integral representations of $\zeta$ numbers. 

One would certainly  like to see at  integrating  $\int_0^{\infty}u\,K_0(u)^n\,\rmd u$ further up to possibly one integration  left on an integrand which might contain, as in (\ref{zetaint}),  a term like $(\log{1+x\over 1-x})^{n-1}$ times a rational function of $x$ yet to be determined, possibly allowing for a derivation of (\ref{final}), and, more generally, of relations between Bessel integrals and $\zeta$ (and polyzeta) numbers.

Finally, the fact that Bessel integrals fall in the category of periods might also be an indication of a deeper meaning yet to be  understood.
  
  Acknowledgments: I acknowledge some useful conversations with  Stefan Mashkevich in particular for helping in the numerics involved in the PSLQ searches of Section (\ref{rewriting}). I have also benefited from numerous technical and helpful discussions  with Jean Desbois.
\pagebreak

\section*{ Appendix : More on integrating  }

\small{

{$\bf n=3$}

From (\ref{nicenice}) 
 \begin{align}\int_{0}^{\infty}\rmd u\,  u
K_{0}(u)^3&= \int_{0}^{1}{\rmd x_1}\log{1+x_1\over 1-x_1} 
{1\over 1+3x_1^2}= \int_{0}^{1}{\rmd x_1}\log{1+x_1\over 1-x_1} 
{2\over (1+x_1)^3(1+({1-x_1\over 1+x_1})^3)}\nonumber\\
&=-{1\over 2}
\int_{0}^{1}{\rmd u}{1+u\over 1+u^3}\log{u}\nonumber\\
&=-{1\over 2}
\int_{0}^{1}{\rmd u}(1+u)\sum_{k=0}^{\infty}(-1)^ku^{3k}\log{u} \nonumber\\
&={1\over 2}\sum_{k=0}^{\infty}(-1)^k({1\over(3k+1)^2}+{1\over(3k+2)^2})={\psi_1(1/3)-\psi_1(2/3)\over 12}\label{moncoco} 
\end{align}
where one has made the change of variable $u={1-x_1\over 1+x_1}$ and used \be\int_0^1\rmd u\, u^k(\log u)^n =(-1)^n{n!\over (k+1)^n}\ee 
In (\ref{moncoco}) $\psi_1(z)$ stands for the polygamma function such that $8\zeta(2)=\psi_1(1/3)+\psi_1(2/3)$, thus relating $\int_{0}^{\infty}\rmd u\,  u
K_{0}(u)^3$ to $\zeta(2)$.

From(\ref{OKKK})
\be\int_{0}^{\infty}\rmd u\,  u
I_0(u)K_{0}(u)^3={1\over 4}\int_0^1\rmd x_1\int_0^{1-x_1}\rmd x_2 {1\over x_1x_2(x_1+x_2)-(x_1+x_2)(1-x_1-x_2)}\ee
With the change of variables $x=x_1+x_2$ and $y=x_1-x_2$ one gets
\begin{align}\int_{0}^{\infty}\rmd u\,  u
I_0(u)K_{0}(u)^3=&{1\over 4}\int_0^1\rmd x\int_0^{x}\rmd y{4\over x}{1\over x-2+y}{1\over x-2-y}\nonumber\\
=&{1\over 2}\int_0^1\rmd x{1\over x(x-2)}\int_0^{x}\rmd y{1\over x-2+y}+{1\over x-2-y}\nonumber\\
=&{1\over 2}\int_0^1\rmd x{1\over x(x-2)}\log(1-x)\nonumber\\
=&-{1\over 2}\int_0^1\rmd u{1\over 1-u^2}\log u={3\zeta(2)\over  8}\end{align}
where one has set $u=1-x$.

{$\bf n=4$}

From (\ref{nicenice})
\begin{align}\int_{0}^{\infty}\rmd u\,  u
K_{0}(u)^4&={1\over 2^{3}} \int_{0}^{1}{\rmd x_1}\int_{0}^{1-x_1}{\rmd x_2}\log{1+x_1+x_2\over 1-x_1-x_2} {1\over x_1+x_2}
{4\over 1-(x_1+x_2)^2+4 x_1x_2}\nonumber\\ 
&={1\over 2^{3}}{1\over 2} \int_{0}^{1}{\rmd x}{1\over x}\log{1+x\over 1-x}\int_{-x}^{x}{\rmd y} 
{4\over 1-y^2}\nonumber\\
&={1\over 2} \int_{0}^{1}{\rmd x}{1\over x}\log{1+x\over 1-x}\int_0^{x}{\rmd y} 
{1\over 1-y^2}\nonumber\\
&={1\over 2}{1\over 2} \int_{0}^{1}{\rmd x}{1\over x}(\log{1+x\over 1-x})^2\nonumber\\
&={1\over 2}\int_{0}^{1}{\rmd u}{1\over 1-u^2}(\log u)^2\nonumber\\
&={1\over 2}\int_{0}^{1}{\rmd u}(\log u)^2\sum_{n=0}^{\infty}u^{2n}={7\zeta(3)\over 8}\label{cb}\end{align}
with the change of variable $u={1-x\over 1+x}$.
 
As a remark one knows (see \cite{moi})  that $\int_{0}^{\infty}\rmd u\,  u^3K_{0}(u)^4$ is a linear combination with rational coefficients of $1$ and $\int_{0}^{\infty}\rmd u\,  u K_{0}(u)^4$ namely
\be 4\int_{0}^{\infty}\rmd u\,  u
K_{0}(u)^4-16\int_{0}^{\infty}\rmd u\,  u^3
K_{0}(u)^4=3\label{100}\ee
From (\ref{cela}) 
\begin{align}\int_{0}^{\infty}\rmd u\,  u^3
K_{0}(u)^4=&\int_{0}^{1}{\rmd x_1} \int_{0}^{1-x_1}{\rmd x_2}\left({1+(x_1+x_2)^2\over 2(x_1+x_2)}\log{1+x_1+x_2\over 1-(x_1+x_2)}-1\right) {1-(x_1+x_2)^2\over (x_1+x_2)^2 }\nonumber\\
&{4x_1x_{2}\over \bigg( 1-(x_1+x_2)^2+4x_1x_{2}\bigg)^2}\label{101}\end{align}
so  that (\ref{100})  implies that
\begin{align}\label{relation} & \int_{0}^{1}{\rmd x_1} \int_{0}^{1-x_1}{\rmd x_2} \log{1+x_1+x_2\over 1-x_1-x_2}{1\over x_1+x_2} 
{2\over1-(x_1+x_2)^2+ 4 x_1x_2}\nonumber\\ 
&-8
\left({1+(x_1+x_2)^2\over 2(x_1+x_2)}\log{1+x_1+x_2\over 1-(x_1+x_2)}-1\right){1-(x_1+x_2)^2\over (x_1+x_2)^2 } 
{4x_1x_{2}\over \bigg(1-(x_1+x_2)^2+4 x_1x_{2}\bigg)^2}=3\end{align}
has to be  satisfied.

With the change of variables $x=x_1+x_2$ and $y=x_1-x_2$  the right hand side of (\ref{101})  becomes
\begin{align} 
&\int_{0}^{1}{\rmd x}\left({1+x^2\over 2x}\log{1+x\over 1-x}-1\right){1-x^2\over x^2}{1\over 2}\left(-x+(1+x^2){1\over 2}\log{1+x\over 1-x}\right)\nonumber\\
&=\int_{0}^{1}{\rmd x}{1-x^2\over 2x}\left({1+x^2\over 2x}\log{1+x\over 1-x}-1\right)^2\end{align}
Finally  (\ref{relation}), becomes 
\be\int_0^1{\rmd x}\left({1\over x}(\log{ 1+x\over 1 - x})^2-4{1-x^2\over x} ({1+x^2\over 2x} \log{ 1+x\over 1 - x} - 1)^2 \right)=3\ee
which is indeed true.

  \pagebreak

\end{document}